\documentclass[%
 reprint,
 amsmath,amssymb,
 aps,
prb,
]{revtex4-2}
\usepackage{hyperref}
\usepackage{graphicx}% Include figure files
\usepackage{dcolumn}% Align table columns on decimal point
\usepackage{bm}% bold math
\begin{document}

\preprint{APS/123-QED}

\title{Solid Solubility in Metallic Hydrogen.}% Force line breaks with \\

\author{Jakkapat Seeyangnok$^{1}$}
 \email{jakkapatjtp@gmail.com} 
 %\altaffiliation[Also at ]{Department of Physics, Faculty of Science, Chulalongkorn University, Bangkok, Thailand.}%Lines break automatically or can be forced with \\
\author{Udomsilp Pinsook$^{1}$}%
 \email{Udomsilp.P@Chula.ac.th}
%\affiliation{Department of Physics, Faculty of Science, Chulalongkorn University, Bangkok, Thailand.}%

\author{Graeme J Ackland$^{2}$}
 \email{gjackland@ed.ac.uk} 
\affiliation{$^{1}$Department of Physics, Faculty of Science, Chulalongkorn University, Bangkok, Thailand.\\$^{2}$Centre for Science at Extreme Conditions, School of Physics and Astronomy, University of Edinburgh, Edinburgh, United Kingdom.}%


\date{\today}% It is always \today, today,

\begin{abstract}Hydrogen  in  its metallic form is the most common material in our solar system, found under the extreme pressure and temperature conditions found in giant planets. Such conditions are inaccessible to experiment and consequently, theoretical work has typically led experiment.  Many remarkable properties are proposed for metallic hydrogen, which is expected to exist in solid form down to low temperatures. Relevant here is that solid metallic hydrogen is not expected to have a close-packed structure, even at extreme pressures. The predicted crystal structure has \cite{pickard2007structure,mcmahon2011ground,mcmahon2012properties,geng2012high,geng2015lattice,geng2019thermodynamic}  a packing fraction of only 0.44 compared with 0.74 for fcc.  Alloying with other metallic elements  can form compounds with atomic hydrogen and much reduced metallization pressure \cite{zurek2009little}.   Here, we investigate the possible solid solubility of a representative range of elements (Be, B, Mg, S, Fe, La) in metallic atomic hydrogen near 500GPa. Zero point energy is highly significant in all cases.  The metallic elements Be, Mg, Fe, and La have a well-defined clathrate-style first neighbour shell at 300K, while the non-metallic B and S have more open arrangements.  
All the elements considered are soluble, in sites substituting for more than one hydrogen atom.  The solid solubility of such a wide range of ambient bonding types: simple metals, covalent materials, transition metals, lanthanides, which gives a strong indication that most elements will dissolve in solid metallic hydrogen.
\end{abstract}

\maketitle

%\tableofcontents
    %\subsection{Introduction.}
    \emph{Introduction -}
    Theory has always led experiment in studies of metallic hydrogen.
    Almost 100 years ago, J.D. Bernal proposed that all materials would become metallic when they are subjected to enough high pressure. 90 years ago, Wigner and Huntington first calculated properties of metallic hydrogen using free electron theory and an assumed bcc structure, predicting a density remarkably close to current estimates \cite{wigner1935possibility}. More recently, with modern techniques, the ground state structures of atomic metallic hydrogen have been theoretically determined as a function of pressure \cite{mcmahon2011ground,mcmahon2012properties,nagao1997structures,natoli1993crystal, barbee1991theoretical,barbee1989theory,ebina1989anisotropic, nagara1989anisotropic,ceperley1987ground,straus1977self,kagan1977equation,brovman1972properties} together with still controversial experimental investigations, as reviewed in \cite{silvera2021phases,dias2017observation,eremets2011conductive,loubeyre2020synchrotron}. 

   %\subsection{Intro - Solubility}
    The behaviour of atomic metallic hydrogen at modest temperatures is also of interest for understanding the structure of giant planets \cite{baraffe2009physical}, black and brown dwarfs, and predicted properties such as high-$T_{c}$ BCS superconductivity \cite{ashcroft1968metallic,mcmahon2011high,zhang2020pseudopotential,zhuang2020evolution},  and a metallic fluid at low or possibly zero temperature \cite{bonev2004quantum}. 

    %\subsection{Intro - Ground-state structures}
Below 500GPa,  many theoretical calculations \cite{mcmahon2011ground,pickard2007structure,johnson2000structure} predict that the most stable structures are molecular phases, some of which are predicted to become metallic (e.g. $Cmca$). At about 500GPa, it is believed that hydrogen molecules dissociate into a monoatomic body-centered tetragonal structure of the $I4_{1}/amd$ space group with a ratio $c/a>1$ \cite{mcmahon2011ground,pickard2007structure} and an isostructural  metastable $I4_{1}/amd$ state with  $c/a<1$, which becomes much less stable with increasing pressure, while the high c/a version remains to 2.5TPa.    Above 2.5TPa, denser packing dominates and there are four possible candidate structures: face-centered orthorhombic with space group $Fmmm$, a hexagonal planar structure of $P6_{3}/mmc$, hexagonal and planar of $R\bar{3}m$, and $R3m$ that range from the least stable to the most stable, respectively.

A surprising feature of this is that $I4_{1}/amd$ has a packing efficiency of only 0.44, very far from the close-packed structures envisioned by early workers (we echo Wigner's observation that bcc is close packed in reciprocal space, which is relevant for free electron materials). The high and low c/a  structures are also observed in cesium-IV and $\beta$-tin, respectively, and the structure can be considered to be stabilized by strong Fermi surface Brillouin zone interactions rather than packing of atoms \cite{ackland2004origin}.  Curiously, although the atoms are not close packed,  $I4_{1}/amd$ is a stable structure of the antiferromagnetic Ising model on an fcc lattice\cite{ackland2023phase}, the analogy being protons and electrons in place of spin-up and spin-down.

In 1968, Ashcroft predicted that high-pressure hydrogen could exhibit superconductivity with a high superconducting transition temperature $T_{c}$ \cite{ashcroft1968metallic}. Consequently, extensive research has been conducted on high-temperature superconductivity in pure atomic metallic hydrogen \cite{dias2017observation,mcminis2015molecular,azadi2014dissociation,mcmahon2011high,zhang2020pseudopotential,zhuang2020evolution}. However, the addition of other elements to hydrogen can lower the metallization pressure by hundreds of GPa by disrupting the covalent bonds forming H$^-$ and even H$_3^-$\cite{zurek2009little,marques2023h2}. Hydrogen-rich compounds demonstrating superconductivity have been synthesized in situ using diamond anvil cells, achieved by elevating pressure in an environment abundant with hydrogen, \cite{eremets2022high}. Notably, hydrogen sulfide was the breakthrough discovery of such high-Tc hydride superconductors\cite{drozdov2015conventional}; Experimental characterization revealed a cubic structure, while  density functional calculations identified its composition as H$_3$S, exhibiting conventional (BCS) superconductivity \cite{duan2014pressure,li2014metallization,drozdov2015conventional,errea2015high}. Recent investigations have unveiled multiple occurrences of high-temperature superconductivity in hydrogen-rich compounds under elevated pressure conditions, thus enabling the synthesis of hydrogen-rich compounds in experimental settings \cite{wang2012superconductive,drozdov2015conventional,peng2017hydrogen,liu2017potential,troyan2021anomalous,kong2021superconductivity,snider2021synthesis,chen2021high,somayazulu2019evidence,drozdov2019superconductivity,semenok2021superconductivity,ma2021high,PinsookReview,van2023competition}. Currently, the experimental pinnacle of $T_c$ is  250-260 K in a lanthanum hydride, thought to be LaH$_{10}$ \cite{drozdov2019superconductivity,somayazulu2019evidence}.
    
    At higher temperatures, the solubility of additional elements  in hydrogen plays a very important role in many properties in the evolution of the interior structure of the planets \cite{wahl2013solubility}, e.g. the surprisingly small core in Jupiter \cite{militzer2022juno,helled2022revelations} which was not previously expected \cite{militzer2008massive}.  
    
    A new generation of structure search codes has enabled theorists to search for new crystal structures generating convex hulls for alloys \cite{pickard2011ab,wang2012calypso}. In such work, the limiting case of solid solubility is often ignored, with end members assumed to be pure elements competing with pure compounds.  In part, this is because 
    the hull is implicitly at 0K, so entropy of solution is ignored; furthermore disordered solid solubility cannot be detected by crystallography, whereas ordered compounds can, even when the atomic concentration of the non-hydrogen element is less than 5\% such as  IH$_{27}$\cite{ackland2018icosahedral,binns2018formation}.  We also note that solid solutions may have unfavorable enthalpy, compensated for by the entropy of mixing. 
    
    In this letter, we investigate the solid solubility of various elements in  metallic atomic hydrogen near 500GPa. 
    We apply density functional theory calculations to compare the free energy of hydrogen with and without each alloying element. An unusual feature is that, because hydrogen is small, the solutes typically replace more than one hydrogen atom in the most stable configurations \cite{hepburn2015transition,ackland2023phase}.

    %%%%%%%%%%%%%%%%%%%%%%%%%%%%%%%%%%%%%%%%%%%%%%%%%%%%%%
    %\subsection{$I4_{1}/amd$ metallic hydrogen and alloys.}
    
    %\begin{figure*}
	%\includegraphics[width=17.5cm]{Figure1.jpg}  
	%\caption{The figures show the radial distribution function (RDF) and the mean square displacement (MSD) of the substitutional alloys. (a-h) show the RDF and the MSD of $4\times4\times2$ supercell of 128 hydrogen atoms with additional alloys consisting of Be, B, Mg, and S for different number of removing hydrogens. (i-l) display the RDF and the MSD of $5\times5\times2$ supercell of 200 hydrogen atoms with additional alloys (a brown ball) consisting of Fe, and La for different number of removing hydrogens.}
	%\label{fig:MSD-RDF-all}
	%\end{figure*}

     \begin{figure}[h]
    \centering
	\includegraphics[width=8.5cm]{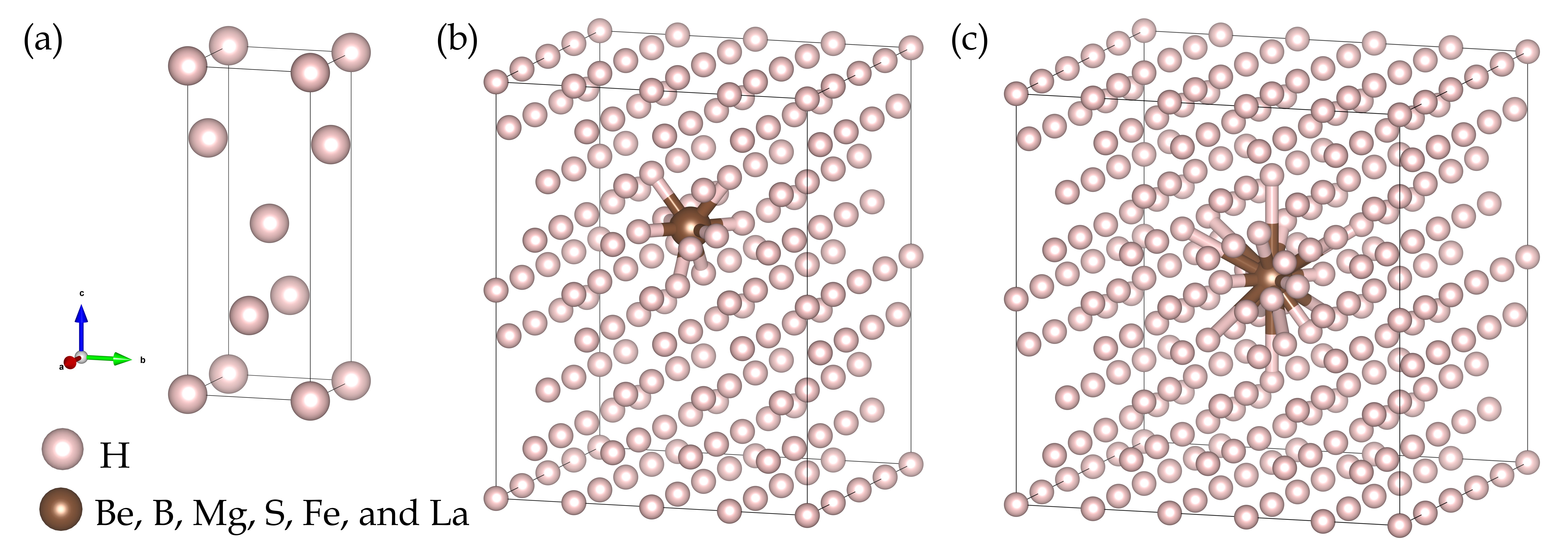}
	\caption{(a) a single cell of the mono-atomic body-centered tetragonal of the $I4_{1}/amd$ space group. (b) $4\times4\times2$ supercell of 128 hydrogen atoms with the substitutional alloy (a brown ball) consisting of Be, B, Mg, or S. (c) $5\times5\times2$ supercell of 200 hydrogen atoms with an additional Fe, or La.}
	\label{fig:i41amd}
	\end{figure}

The calculations use Born-Oppenheimer molecular dynamics (BOMD) implemented in the Cambridge Serial Total Energy Package (CASTEP) \cite{CASTEP} within the NPT ensemble at 300 K. Detailed computational methods are provided in the supplementary material.
 
    \emph{Reference States -}  We consider solution of Be, B, Mg, S, Fe, and La in $I4_{1}/amd$ metallic hydrogen with at 500 GPa and 300 K.
    The solute  reference states under these conditions are recalculated using structures based on previous work as follows: For Beryllium, a body-centered cubic (BCC) structure within the $Im\overline{3}m$ space group, as reported in \cite{wu2021high}. Boron, according to \cite{oganov2011boron}, exhibits an orthorhombic $\alpha$-Ga structure belonging to the $Cmce$ space group, with four atoms in the primitive unit cell above 90 GPa. This structure is believed to persist up to 500 GPa and possibly exhibit supersolid properties.  Magnesium is identified as BCC, consistent with \cite{stinton2014equation,beason2021shock}. Similarly, Sulfur and Iron adopt a hexagonal closed-packed structure (HCP) in the $P6_3/mmc$ space group \cite{whaley2020first,pickard2009stable}. Finally, lanthanum is described by the distorted FCC in the $I4/mmm$ space group\cite{chen2022phase}. Using Density Functional Perturbation Theory (DFPT) and BOMD, we find that these reference structures maintain dynamical and thermal stability at both zero temperature and at room temperature (300 K) under the extreme pressure of 500 GPa.

\emph{Phonon free energy -} we use BOMD as the basis for free-energy and zero-point calculations. Specifically, we quantize the phonon density of states (PhDOS) obtained from velocity autocorrelation \cite{pinsook1,pinsook2}.  For pure elements, this was compared  with the DFPT quasiharmonic lattice dynamics calculations: good agreement was found, validating the BOMD approach and demonstrating the lack of  anharmonicity in those cases. Quantum effects and anharmonicity are important in hydrogen systems, and zero-point energy contributes significantly to both the calculated energy and pressure \cite{van2023competition,ackland2015appraisal}. When we calculate the PhDOS of $I4_{1}/amd$ atomic hydrogen using BOMD, this anharmonicity is inherently accounted for by finite displacements%: BOMD gives a different PhDOS from DFPT, which is computed based on the harmonic approximation using the potential curvature at the equilibrium position. By quantising the measured phonon frequencies, hydrogen can be treated as a quantum atom. 
    %\subsection{Structure of substitutional alloy}
     
    \begin{figure*}
	\includegraphics[width=18cm]{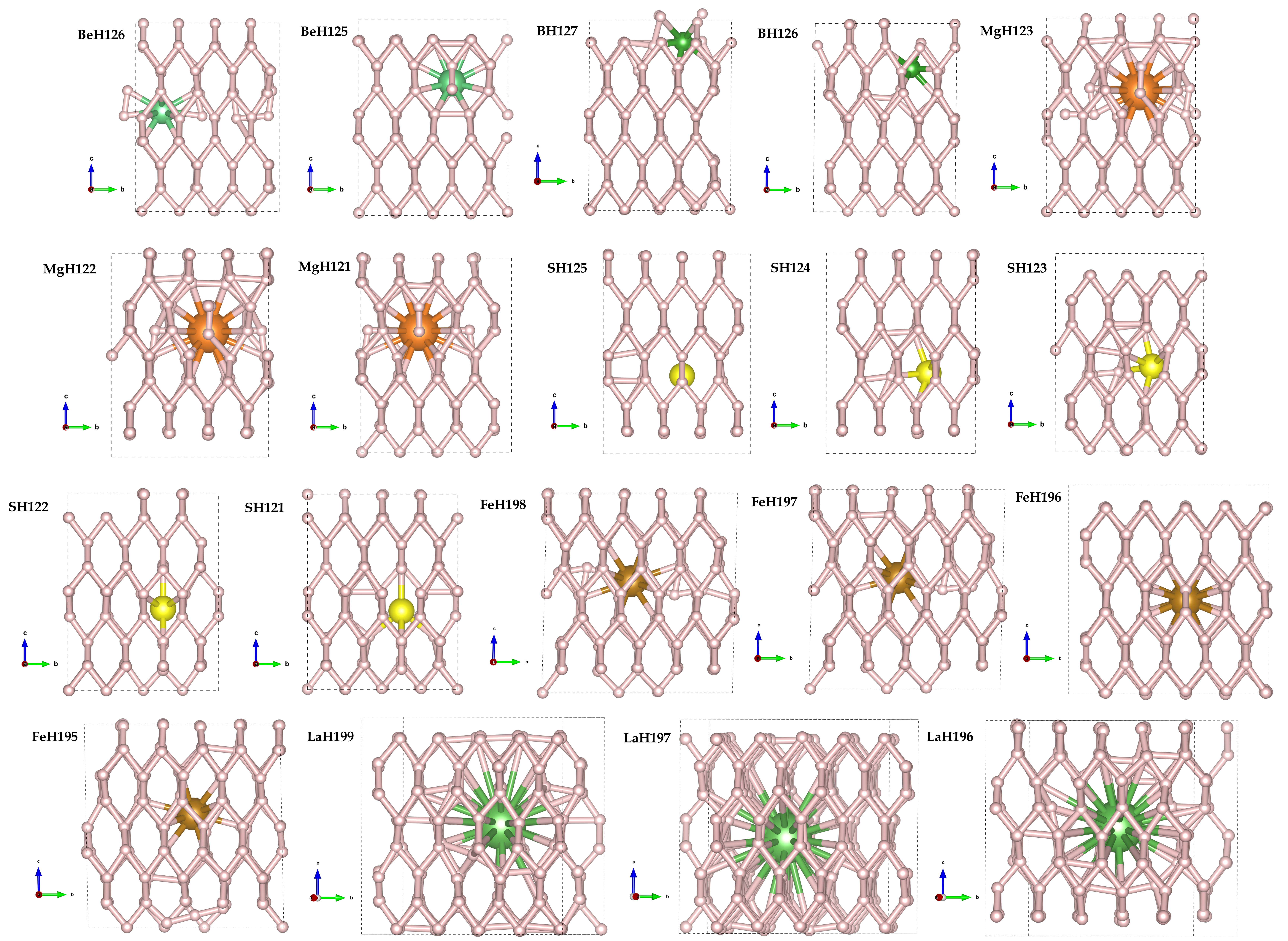}
	\caption{The figures show optimization of structures of substitutional alloys at of the final snapshot of BOMD with the selected concentrations with the sharpest peak structures in the RDF along with minimal MSD. }
	\label{fig:Structure-substitutional-alloy}
	\end{figure*}
 
     %\textbf{\emph{Structures of substitutional alloys.}} Although "atomic size" is not a very rigorously defined concept at 500 GPa, we can expect hydrogen to  be small, and that solid substitution may involve replacing more than one hydrogen with a single solute atom. Consequently, we repeated the BOMD solubility calculations by replacing  different numbers  of hydrogens and inserting an alloying element. Fig~\ref{fig:MSD-RDF-all} shows the radial distribution function (RDF) and the mean square displacement (MSD) of the BOMD-NPT ensemble averaged over 10.0 ps with up to nine hydrogen atoms per impurity removed.  Most of the structures of substitutional alloys have H-H RDFs structures similar to $I4_1 /amd$, as shown in  Fig~\ref{fig:MSD-RDF-all}(a, c, e, g, i, k). This indicates that the alloys are indeed substitutional and do not induce a structural transformation in the original $I4_1 /amd$ of metallic hydrogen.  However, in some cases there is a complete rearrangement of the atoms. 

     \emph{Structures of substitutional alloys -} Although "atomic size" is not a very rigorously defined concept at 500 GPa, we can expect hydrogen to be small, because it is, essentially, just a proton.  Therefore we consider that solid substitution may involve replacing more than one hydrogen with a single solute atom. Consequently, we set up BOMD solubility calculations by replacing  different numbers  of hydrogens and inserting an alloying element. We monitor the mean squared displacement (MSD) of the atoms to determine dynamical stability. After equilibration, most of the supercells containing  alloys have stable MSD and H-H RDFs structures similar to $I4_1 /amd$. This indicates that the alloys are indeed substitutional and no structural transformation in the original $I4_1 /amd$ of metallic hydrogen has occurred.  However, in some cases there is a complete rearrangement of the atoms throughout the cell, typically associated with high enthalpy. 
     
     We found that calculations based on 128-hydrogen atom supercells were large enough to obtain $I4_1 /amd$-like RDFs for the hydrogen.   
    
    The suitable number of hydrogens in the substitutional alloy system can also be observed in the MSD of the system as shown in the supplementary material. %, as shown in (b, d, f, h, j, l) of Fig~\ref{fig:MSD-RDF-all}. 
    It is evident that the stable structures exhibit sharp peaks in the RDF corresponding to well defined  hydrogen positions, and the MSD tends to be small and stable.  Alternately, if too many hydrogens are removed, vacancies are created which diffuse throughout the system, leading to a constant increase in the MSD. Equivalently, removing too few hydrogens causes self-interstitial hydrogen defects to be emitted, which also diffuse and give a steady increase in the MSD.
     In reality, excess hydrogens or hydrogen vacancies could migrate to the surface of the system, but in the simulation, they re-enter the system due to periodic boundary conditions. In extremely unstable cases, (e.g.  MgH$_{128}$ and MgH$_{127}$) we find possible melted structures with all atoms diffusing. 
    Stable solute structures with constant MSD occur for the supercells shown in Table\ref{tab:freeenergy}, and we use only these for our detailed  calculations of free energy of solution $g_{\text{sol}}$.

\begin{table}[h]%[b] The best place to locate the table environment is directly after its first reference in text
\caption{\label{tab:freeenergy}
Analysed supercells with number of removed hydrogens per solute, and associated ensemble calculation of the change (mixed - unmixed) in enthalpy (eV/impurity), entropy (eV/impurity) ($S_{vibra}+S_{conf.}$), zero point energy (eV/impurity) and the free energy of solution (eV/unit cell) for different impurities substituted for various numbers of hydrogens  (-H) from BOMD with the NPT ensemble at 300K and 500GPa. Full details of the uncertainty calculation are given in Supplementary Materials}
\begin{ruledtabular}
\begin{tabular}{cccccccc}
\textrm{Supercell}&
\textrm{-H}&
\textrm{$\Delta H$}&
\textrm{$-T\Delta S$}&
\textrm{$\Delta U_{ZPE}$}&
\textrm{$g_{sol}$}&\\
\colrule
            BeH$_{126}$ &2& -0.42$\pm$0.09  & -0.54$\pm$0.05 & -1.20$\pm$0.87 & -2.15$\pm$0.88\\
            BeH$_{125}$ &3& -0.49$\pm$0.10  & -0.43$\pm$0.08 & -0.55$\pm$0.69 & -1.48$\pm$0.71 \\
            \hline            
            BH$_{127}$ &1& 1.59$\pm$0.11 & -0.47$\pm$0.02 & -1.44$\pm$0.86 & -0.32$\pm$0.87 \\
            BH$_{126}$ &2& 1.52$\pm$0.11  & -0.45$\pm$0.04 & -1.620$\pm$0.87 & -0.55$\pm$0.88 \\
            BH$_{125}$ &3& 1.35$\pm$0.12  & -0.41$\pm$0.06 & -0.46$\pm$0.87 & 0.48$\pm$0.89 \\
            \hline
            MgH$_{123}$ &5& -6.86$\pm$0.09  & -0.42$\pm$0.04 & 0.020$\pm$0.96 & -7.26$\pm$0.97 \\
            MgH$_{122}$ &6& -6.88$\pm$0.13  & -0.37$\pm$0.05 & 1.38$\pm$0.72 & -5.87$\pm$0.74 \\
            MgH$_{121}$ &7& -7.13$\pm$0.11  & -0.36$\pm$0.06 & 0.40$\pm$1.01 & -7.09$\pm$1.02 \\
            \hline
            SH$_{125}$ &3& 1.32$\pm$0.11 & -0.56$\pm$0.08 & -1.57$\pm$0.81 & -0.81$\pm$0.83 \\
            SH$_{124}$ &4& 1.25$\pm$0.12 & -0.63$\pm$0.08 & -2.31$\pm$1.01 & -1.69$\pm$1.02 \\
            SH$_{123}$ &5& 1.08$\pm$0.09 & -0.49$\pm$0.06 & -1.78$\pm$1.08 & -1.19$\pm$1.09 \\
            SH$_{122}$ &6& 1.06$\pm$0.09 & -0.48$\pm$0.03 & -0.73$\pm$0.87 & -0.16$\pm$0.88 \\
            SH$_{121}$ &7& 0.86$\pm$0.12 & -0.54$\pm$0.06 & -0.97$\pm$0.63 & -0.66$\pm$0.65 \\
            \hline
            FeH$_{198}$ &2& -0.83$\pm$0.13 & -0.58$\pm$0.04 & 0.09$\pm$1.07 & -1.31$\pm$1.09 \\  
            FeH$_{197}$ &3& -0.78$\pm$0.14 & -0.48$\pm$0.11 & -0.60$\pm$1.74 & -1.86$\pm$1.76 \\  
            FeH$_{196}$ &4& -0.97$\pm$0.18 & -0.51$\pm$0.05 & -0.21$\pm$1.49 & -1.68$\pm$1.51 \\  
            FeH$_{195}$ &5& -0.80$\pm$0.13 & -0.55$\pm$0.05 & -0.87$\pm$1.47 & -2.22$\pm$1.49 \\  
            \hline
            LaH$_{199}$ &1& -1.03$\pm$0.22 & -0.75$\pm$0.14 & -1.03$\pm$1.37 & -2.81$\pm$1.41 \\
            LaH$_{197}$ &3& -1.24$\pm$0.13 & -0.69$\pm$0.17 & -1.13$\pm$1.57 & -3.06$\pm$1.59 \\
            LaH$_{196}$ &4& -1.07$\pm$0.14 & -0.83$\pm$0.22 & -1.39$\pm$1.68 & -3.28$\pm$1.71 
            \\
\end{tabular}
\end{ruledtabular}
\end{table} 

    %\subsection{The solubility}
    \emph{The solubility -}  Table~\ref{tab:freeenergy} presents the calculated free energies for the most promising supercells as discussed previously. This includes the mean enthalpy and the entropy, which encompasses both the vibrational entropy evaluated using the quasi-harmonic approximation and the  configurational entropy at the concentration of the supercell.  To emphasize the importance of zero-point energy, this is listed separately.  We find that substitutional alloys with metals (Be, Mg, Fe, La) exhibit enthalpies $\Delta H$ that favor mixing, whereas B and S have positive enthalpy of solution. However, once the zero-point energy $\Delta U_{\text{ZPE}}$ and the entropy $-T\Delta S$ are included, all elements have at least one arrangement with a negative free energy of solution $g_{\text{sol}}$.  This demonstrates that Be, B, S, Fe, and La can dissolve in solid $I4_1/amd$ metallic hydrogen at 500 GPa and room temperature.

    \begin{figure}[h]
    \centering
	\includegraphics[width=8.5cm]{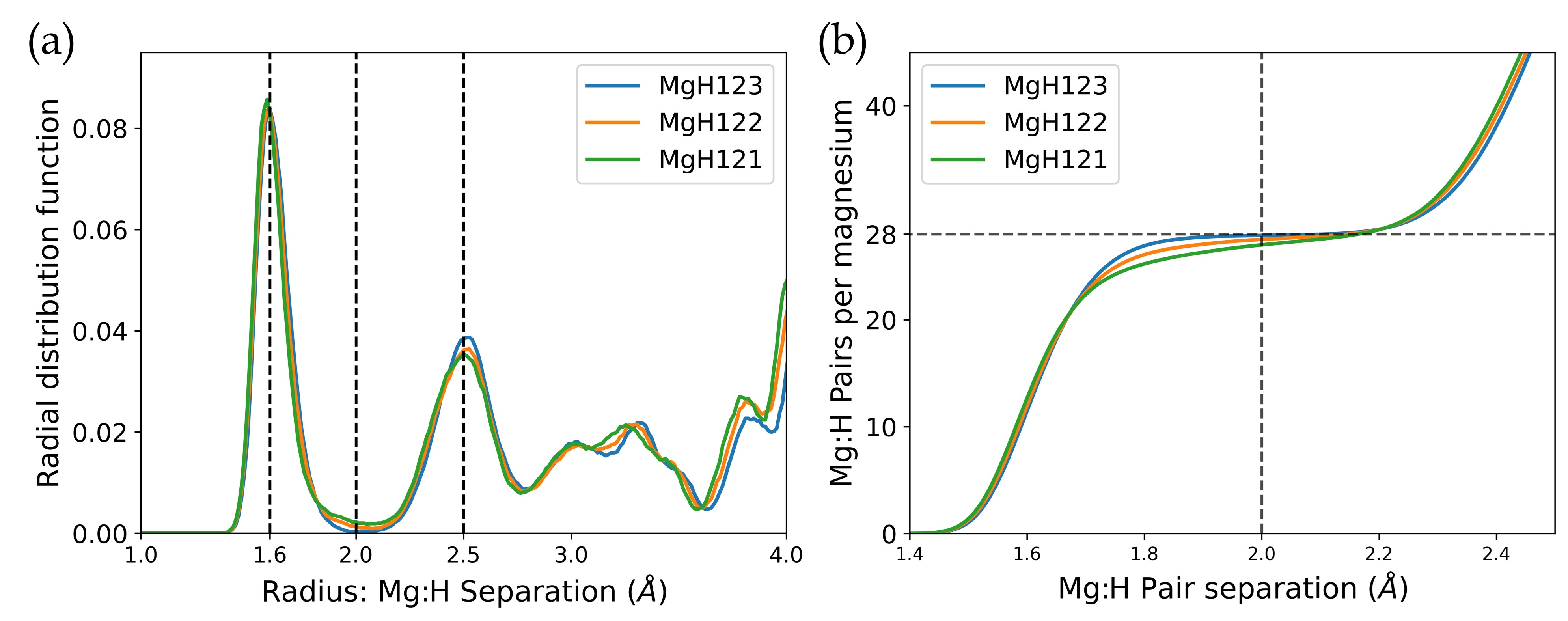}
	\caption{(a) The radial distribution function between Mg and hydrogens. (b) The cumulative distribution function (RDF) of Mg:H pairs per Mg: the large flat region demonstrates the well defined 28-fold coordination shell.}
	\label{fig:mgh-pairs}
	\end{figure}

\begin{figure}[h]
    \centering
	\includegraphics[width=8.5cm]{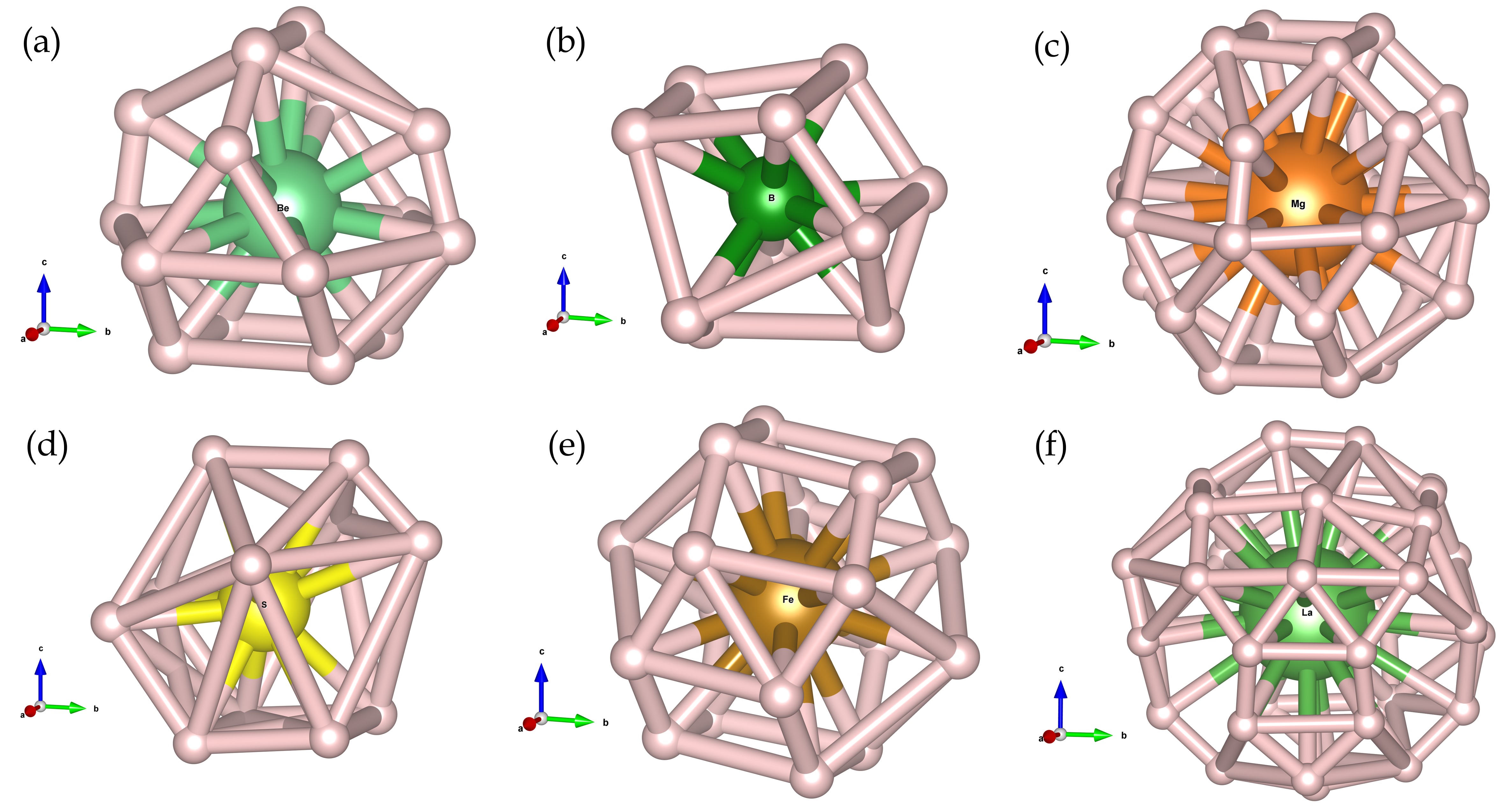}
	\caption{(a-f)  snapshots of the X-H coordination shell (X=Alloys) from the BOMD,  with only the near-neighbor shell of hydrogen for each impurity  BeH$_{16}$, BH$_{10}$, MgH$_{28}$, SH$_{11}$, FeH$_{18}$, and LaH$_{36}$, respectively. The radial cutoff between alloys and hydrogens are 1.65 \AA, 1.50 \AA, 2.00 \AA, 1.65 \AA, 1.75 \AA, and 2.25 \AA ~ for Be, Mg, S, Fe, and La, respectively.}
	\label{fig:coordination}
	\end{figure}

    The optimized structures are shown in Fig.~\ref{fig:Structure-substitutional-alloy}. We identify two possible types of hydrogen structures surrounding the alloys: clathrate-like and non-clathrate-like structures, similar to H$_3$S, \cite{li2014metallization,duan2014pressure}, and LaH$_{10}$, \cite{drozdov2019superconductivity,somayazulu2019evidence}, respectively. Clathrate-like structures are present around the metal solutes Be, Fe, Mg and La and also for B. In these clathrate-like structures, there is a clear gap between the first  and second neighbour shells, as shown in Fig.~\ref{fig:mgh-pairs}(a) for Mg and in Supplementary material for other elements. Mg-H exhibits a peaked RDF and flat region of the CDF Fig.~\ref{fig:mgh-pairs}(b) which reveals a first coordination shell of twenty-eight hydrogens at a distance around 2.0 \AA.  
    Similar arrangements are found in other metals (fig
    \ref{fig:coordination}) with the coordination roughly following the ionic radii of the solutes. The structure of the X-H (X=Alloys) coordination shells are shown in Figure~\ref{fig:coordination}. We identified the near-neighbor shell of hydrogen for each impurity as BeH$_{16}$, BH$_{10}$, MgH$_{28}$, FeH$_{18}$, and LaH$_{36}$. Interestingly, the number of atoms in the clathrate shell is largely independent of the number of hydrogens removed in the starting configuration. %All details of the X-H RDFs and the associated CDFs can be found in the supplementary material.  

  Again, sulfur is different - the 11 atom shell shown in Figure~\ref{fig:coordination} 
  is typical, but there is no consistency across the calculations with coordinations between 8 and 13 neighbours being observed in differrent snapshots.

    %\begin{table}[b]%The best place to locate the table environment is directly after its first reference in text
%\caption{\label{tab:coordination} DONT UNDERSTAND???? The table shows the X-H coordination of each impurity within their corresponding supercell, along with the radial cutoff between the impurity (X-H cutoff (\AA)) and the near-neighbor shell of hydrogen and the radial cutoff between the near-neighbor shell of hydrogens H-H cutoff (\AA).}
%\begin{ruledtabular}
%\begin{tabular}{cccccccc}
%\textrm{Supercell}&
%\textrm{X-H cutoff (\AA)}&
%\textrm{H-H cutoff (\AA)}&
%\textrm{The coordination}\\
%\colrule
	%	BeH$_{126}$ & 1.65 & 1.50 & BeH$_{16}$  \\
	%	BH$_{126}$ & 1.50 & 1.60 & BH$_{10}$  \\
	%	MgH$_{123}$ & 2.00 & 1.30 & MgH$_{28}$ \\
    %    SH$_{124}$ & 1.65 & 2.00 & SH$_{11}$ \\
	%	FeH$_{195}$ & 1.75 & 1.55 & FeH$_{18}$ \\
	%	LaH$_{196}$ & 2.25 & 1.50 & LaH$_{36}$ \\
%\end{tabular}
%\end{ruledtabular}
%\end{table}   

    %\subsection{Conclusion.}
    \emph{Conclusion -} We investigated the solubility of a range of elements in pressure-metallized hydrogen at 500 GPa. These form substitutional alloys with several hydrogens being removed for each impurity.

 There are characteristic clathrate-type structures around each element with each element having well-defined coordination number, scaling with atomic size.  Curiously, sulfur is an outlier with no well defined coordination.
    %We found that sharp peaks in the radial distribution function (RDF) correspond to a stable arrangement of hydrogens with flat mean square displacements (MSD), indicating no hydrogen diffusion. However, in other cases the diffusion of excess hydrogen atoms and hydrogen vacancies throughout the system leads to a constant increase in the MSD. Such excess hydrogens arise from a poor choice of initial conditions and persist due to periodic boundary conditions.
    From the free energy calculations, the metallic elements Be, Mg, Fe, and La have enthalpies $\Delta H$ favoring  alloying with $I4_1 /amd$.  Non-metallic elements B and S have unfavourable enthalpy of mixing. However, once the zero-point energy $\Delta U_{\text{ZPE}}$ and the entropy $-T\Delta S$, is included, all materials yield negative free energies of solubility $g_{\text{sol}}$ at $\sim 1\%$ concentration.
    This demonstrates that Be, B, Mg S, Fe, and La can dissolve in solid $I4_1 /amd$ metallic hydrogen at 500 GPa and room temperature.
    
    These elements were chosen to cover a wide range of ambient bonding types: simple metals, covalent materials, transition metals, lanthanides, which gives a strong indication that most elements will dissolve in solid metallic hydrogen.  Interestingly, the preference for dissolving metallic elements on metallic hydrogen contrasts with the solubility of non-metallic elements and molecules in molecular hydrogen.

    This work implies that the standard computational methodology of structure-search plus convex hull is inappropriate for most systems, since pure $I4_1/amd$ is not a valid end member.  It also serves to illustrate one reason why making metallic hydrogen experimentally is so  difficult.  Not only are the required pressures near the limit of what is currently achievable with diamond anvil technology, but maintaining sample purity for such a super-solvent is extremely challenging.

\begin{acknowledgments}
	This research project is supported by the Second Century Fund (C2F), Chulalongkorn University. GJA acknowledges funding from the ERC project Hecate. This work used the Cirrus UK National Tier-2 HPC Service at EPCC (http://www.cirrus.ac.uk) funded by the University of Edinburgh and EPSRC (EP/P020267/1). This work also used the ARCHER2 UK National Supercomputing Service (https://www.archer2.ac.uk). We acknowledge the supporting computing infrastructure provided by NSTDA, CU, CUAASC, NSRF via PMUB [B05F650021, B37G660013] (Thailand). URL:www.e-science.in.th. We thank Pattanasak Teeratchanan, David Ceperley and Jeffrey M. McMahon for their valuable suggestions on DFT(QE)-related issues for studying metallic hydrogen. 
\end{acknowledgments}

%\nocite{*}

%\bibliography{apssamp}
%\end{document}

% ****** End of file apssamp.tex ******
\end{document}